\documentclass[nopreprint]{jasatex}

\usepackage{amsmath,amsfonts,epsfig}
\usepackage[latin1]{inputenc}

\usepackage{amssymb, latexsym}
\usepackage{amsmath} 
\usepackage{amsthm}
\usepackage{amsfonts,epsfig}

\usepackage{gensymb}
\usepackage{enumerate}

\usepackage{graphicx}
\usepackage{epstopdf}
\usepackage{caption}
\usepackage[list=true]{subcaption}
\usepackage{subfig}
\begin{document}

\title{A robust and passive method for geometric calibration of large arrays}
%
\author{Charles Vanwynsberghe}
\email{charles.vanwynsberghe@upmc.fr}
\author{Pascal Challande, Jacques Marchal, Régis Marchiano, François Ollivier}
\affiliation{Sorbonne Universités, UPMC Univ Paris 06, CNRS,\\
  UMR 7190 Institut Jean Le Rond d'Alembert,
    F-78210 Saint-Cyr-l'\'Ecole, France}

%
%
%
%
%

\begin{abstract}
This paper presents a complete strategy for the geometry estimation of large microphone arrays of arbitrary shape. Largeness is intended here in both number of microphones (hundreds) and size (few meters). Such arrays can be used for various applications in open or confined spaces like acoustical imaging, source identification, or speech processing. For so large array systems, measuring the geometry by-hand is impractical. Therefore a blind passive method is proposed. It is based on the analysis of the background acoustic noise, supposed to be a diffuse field. The proposed strategy is a two-step process. First the pairwise microphone distances are identified by matching their measured coherence function to the one predicted by the diffuse field theory. Secondly, a robust MultiDimensional Scaling algorithm is adapted and implemented. It takes advantage of local characteristics to reduce the set of distances and infer the geometry of the array. This work is an extension of previous studies and it overcomes unsolved drawbacks. In particular it deals efficiently with the outliers known to ruin standard MDS algorithms. Experimental proofs of this ability are presented by treating the case of two arrays. They show that the proposed improvements manage large spatial arrays. \textit{Copyright (2016) Acoustical Society of America}
\end{abstract}

\pacs{43.60.Fg\\
Preprint version \footnote{This article may be downloaded for personal use only. Any other use requires prior permission of the author and the Acoustical Society of America.}. The following article appeared in (J. Acoust. Soc. Am. 139, 1252) and may be found at\\
\url{http://dx.doi.org/10.1121/1.4944566}}

\maketitle

\section{Introduction}

Microphone array systems are involved in a wide range of applications, such as localization and identification of sources, acoustic imaging, or speech enhancement \cite{Bai2013}. They classically make use of spatial filtering techniques which require an accurate knowledge of the microphones location. An important point which determines the efficiency of these techniques is the \textit{largeness} of the array. It has been proven that a large aperture \textit{i.e.} a wide spatial range \cite{Williams1999,Sachar2001}, as well as a large number of microphones \cite{Bai2013} result in improved performances. Some projects tackled the design and fabrication of large arrays, with 512 \cite{Silverman1998} or 1020 \cite{Weinstein2004} elements.

But so far increasing the number of microphones was tricky because of the heavy hardware required, involving wires and conditioning circuitry for each sensor. The individual cost of a standard acoustic measurement channel also makes it difficult to be multiplied at will. The recent development of numerical microphones based on the MicroElectroMechanical Systems (MEMS) technology is on the verge of revolutionizing acoustic array sensing. Indeed, the most recent MEMS microphone chip solutions embed the signal conditioning circuitry together with the analog-to-digital conversion, and provide a digital output of the measured pressure. It makes possible long range transmissions required by very large arrays of microphones. Together with their numerical recording systems, they are not only affordable but also light and relatively easy to deploy (see \cite{Hafizovic2012}  for 300 elements, \cite{Vanwynsberghe2015} for 128). Moreover the MEMS acoustic sensors present individual and statistical characteristics perfectly suitable for acoustic array sensing techniques since it has been shown that their frequency responses and the directivities are homogenous \cite{Vanwynsberghe2015}. 

Imaging purposes require to know the position of each microphone with a sufficient accuracy. But as the array spatially extends, as well as the number of microphones increases, the measurement of their individual position becomes awkward using standard metrological means, especially for 3D or random geometries. Therefore efficient geometric calibration techniques are required, that should be easy, fast and accurate whatever the application and the array shape. The problem has already been studied, and several methods have been proposed. There are \textit{active} methods requiring external sound sources and \textit{passive} methods analyzing the ambient sound. 

The former methods are mainly based on the estimation of the time difference of arrival between microphones and involve a soundfield containing sources, with \cite{Sachar2005} or without \cite{Crocco2012} \textit{a priori} knowledge of the source signals. These methods are well suited to an accurate positioning of microphones, and are usable for large arrays  \cite{Sachar2005,Khanal2013}. But they have drawbacks: they require external sources and a strict measurement protocol, and their use is limited in a noisy or reverberant environment.

An alternative strategy based on the use of a diffuse field for the calibration has been proposed by McCowan \textit{et al.} \cite{McCowan2008}. They point out that the coherence between two microphones is a function only of the distance between the microphones for a chosen frequency. Thus, by recording in a reverberant environment and by fitting the theoretical and experimental coherences for all the pairs of microphones, they obtained the pairwise microphone distances. To complete the procedure, the coherences are measured several times, providing each time an estimate of the distance for each pair of microphones. In order to reduce the resulting set to one distance per pair of microphones, the data are analyzed by K-means clustering taking the centroid of the cluster as final estimate of the distance. Then, the positions of the individual microphones are derived from the set of pairwise distances by using classic multi-dimensional scaling (MDS). They applied this methodology to real configurations and were able to retrieve the position of 8 microphones with a centimetric precision. Despite a very inspiring idea, this approach suffers several drawbacks. In particular, the measured coherences are very noisy and the fitting procedure provides pairwise distance estimates with a large scattering. The method is not robust due to its sensitivity to wrong input data. This is especially true in the case of large arrays (with large extent or large number of microphones) which are the aim of the presented method. Different improvements have been suggested. Hennecke \textit{et al.} \cite{Hennecke2009} propose a hierarchical approach based on local shape calibration. But their study applies to planar arrays with only 8 microphones. Moreover, the ambient diffuse field requires to be boosted with a controlled white noise generated by a moving loudspeaker. More recently, Taghizadeh \textit{et al.} \cite{Taghizadeh2014} showed that averaging the coherence drastically improves the fitting and strengthens the input to the MDS algorithm. Furthermore, they achieved this task by means of a passive analysis of a natural soundfield. But they also note that coherence fitting can lead to highly erroneous estimates of pairwise distances. In order to enhance the data selection, they successfully implement a standard 2D histogram clustering method, but it is supervised and becomes tricky for large arrays. Finally the main limitation of this method remains in the estimation of pairwise distances when microphones are distant \cite{Taghizadeh2014}. In this case, coherence fitting becomes impractical and is likely to lead to wrong distance estimates (outliers) which cannot be identified straightforwardly in the frame of blind geometric calibration of large arrays. Yet the outliers are unavoidable due to the combination of a great number of pairs and a large spatial extent. These limitations restricted the implementation to planar arrays with small spatial extent (about 25 cm) and a small number of microphones (8).

Reverberant environments constitute common and realistic situations: they create complex soundfields, for which enhanced array applications have been proposed \cite{Ward2003,Nowakowski2015,Chardon2015}. Thus microphone localization is of interest in reverberant rooms, and only needs the prior knowledge of the Schroeder frequency to verify the diffuseness. In this paper, we propose to extend the shape calibration in diffuse field for 2D or 3D microphone arrays with both a large extent (more than $1$~m) and a large number of microphones (more than 100) by introducing improvements to the aforementioned methods. The efficiency of the new shape calibration process is assessed by analyzing its impact on a simulated acoustic imaging scenario (highly dependent on the individual position of microphones). The global methodology follows the seminal studies of McCowan \textit{et al.} \cite{McCowan2008} and Taghizadeh \textit{et al.} \cite{Taghizadeh2014}. Nevertheless, to achieve the goals, important modifications are made at each step to improve robustness.

First, the global methodology is recalled in section \ref{sec:Methodology} following the strategies adopted in previous studies \cite{McCowan2008}, \cite{Hennecke2009}, \cite{Taghizadeh2014}. The improvements brought to the estimation of the pairwise distances is presented in section \ref{sec:EstimationOfThePairWiseMicrophoneDistances}. They are mainly modifications to the coherence function estimator and on the optimization of the spectral analysis parameters. Even with these improvements, the resulting data still contain outliers which are problematic for the MDS methods. To solve the problem a robust MDS process is presented in section \ref{sec:EstimationOfThePositionsOfTheMicrophones}. It has recently been introduced by Forero \textit{et al.} \cite{Forero2012}. They propose a generic Robust MDS (RMDS) method which relies on identifying and inhibiting errors in outliers. However the RMDS input dataset cannot contain more than a few of these outliers; this \textit{sparsity} condition must be guaranteed and should not depend on the largeness of the array. For this purpose the paper derives the Local and Robust MDS (LRMDS): it promotes proximity and disregards the high pairwise distances. The efficiency of this adaptation is asserted experimentally in section \ref{sec:Exps} using two arrays with 2D and 3D shape respectively. The considered approach is fully passive since the measurement protocol is based on the analysis of the naturally diffuse soundfield. The results are evaluated in terms of an overall geometrical error. Finally, in the last section, the suitability of the calibration process is discussed in a simulated scenario with regards to a standard acoustical imaging application.

\section{Methodology}
\label{sec:Methodology}
\subsection{Problem statement}
Let an array of $M$ microphones with unknown individual positions observe an arbitrary acoustic field. The $m$-th microphone records the time-history acoustic pressure $p_m(t)$ whose frequency spectrum is denoted $\hat p_m(f)$. The unknowns of our geometric calibration problem are the positions of the microphones denoted $\mathbf{x}_1$, $\mathbf{x}_2$, ..., $\mathbf{x}_M$. The spatial dimension $D$ of the array though is supposed to be known \textit{a priori} ($D=1$ for linear, $D=2$ for two-dimensional array, and $D=3$ for a full three-dimensional). For notation convenience, we define the unknown matrix $\mathbf{X} = [\mathbf{x}_1,...,\mathbf{x}_M]$ of size $D \times M$, containing all the unknown microphone positions.

\subsection{Process overview}
\begin{figure}[t]
    \begin{center}
        \includegraphics[width=0.8\linewidth]{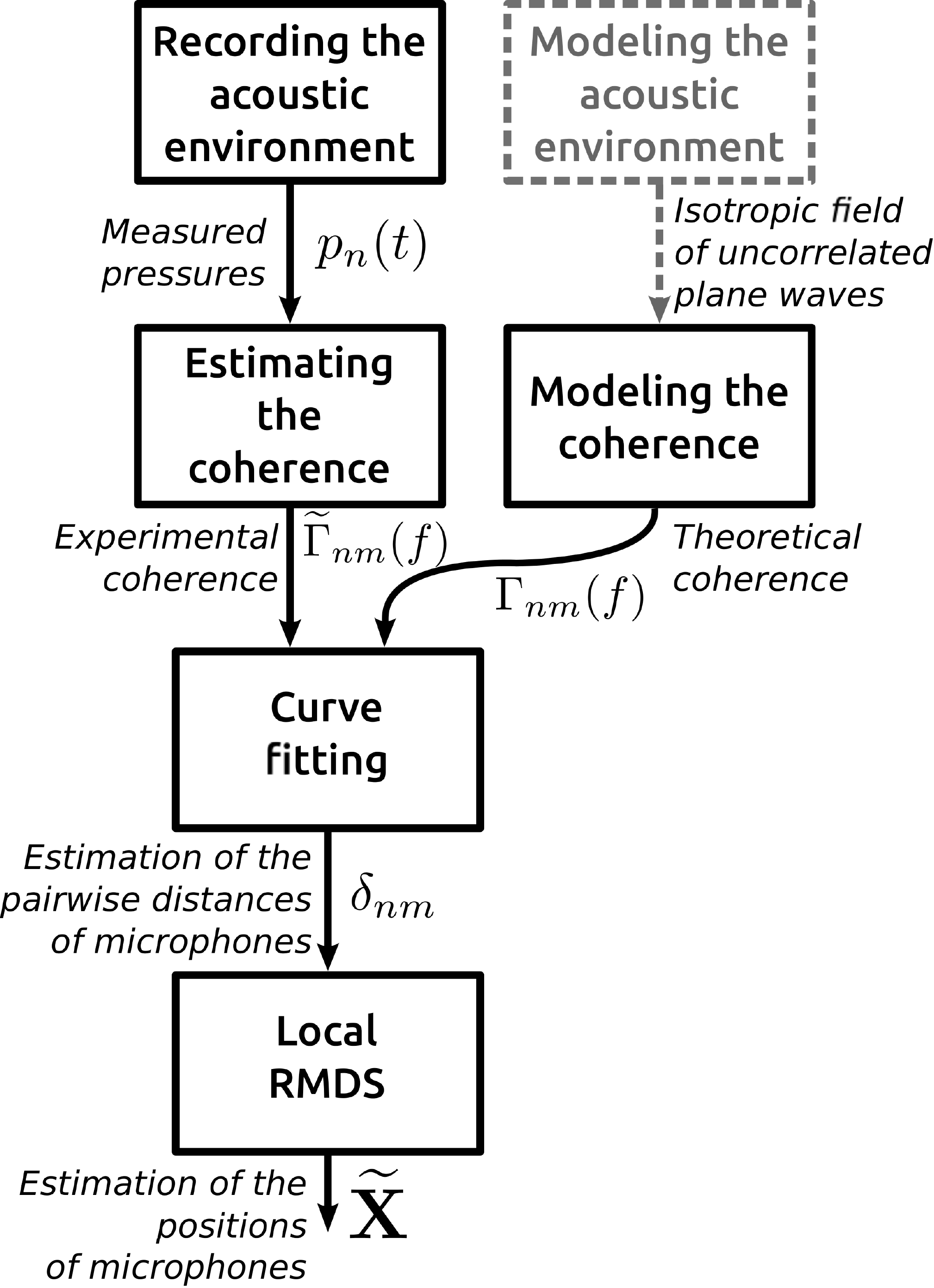}
    \end{center}
    \caption{Functional chart of the geometric calibration process.}
    \label{fig:CalibChart}
\end{figure}

The goal of the proposed method is to find the positions of the microphones. Achieving this goal involves three successive steps.
First, the ambient soundfield is recorded by all the microphones in the array. 
Then, the pairwise microphone distances are estimated from those measurements. This estimation step is a crucial point of the method. It relies on the modeling of the ambient acoustic field as a diffuse field. Indeed in this particular case, the coherence of the field between two locations depends only on the distance. Therefore, the latter can be estimated by fitting the coherence measured with 2 microphones with the theoretical one. The last step derives the microphones individual locations $\mathbf{\widetilde X} = [\mathbf{\widetilde x}_1,...,\mathbf{\widetilde x}_M]$ by computing the redundant information lying in the pairwise microphone distances dataset. This is performed using an enhanced MDS algorithm. The redundancy of the distances dataset ($\mathcal{O}(M^2)$) comes from the fact that $M-1$ couple combinations exist for each microphone. For large arrays, it is much higher than the number of array elements $(\mathcal{O}(M))$.
The overall process is summarized in figure \ref{fig:CalibChart}. 

\section{Pairwise microphone distance estimation}
\label{sec:EstimationOfThePairWiseMicrophoneDistances}
\subsection{Pairwise coherence Estimation}
\subsubsection{Theoretical model for the coherence in diffuse field}
\label{sec:CoherenceModel}

A well known frequency model exists for the coherence function between pressures measured at two distinct locations in a diffuse field.  This model considers an isotropic distribution of uncorrelated plane waves \cite{Jacobsen2000,Nelisse1997}. We suppose that this model holds over a broad frequency band $f \in \left[f_{\text{min}}; f_{\text{max}}\right]$. By definition, the coherence function between microphones $n$ and $m$ is:
\begin{equation}
    \Gamma_{nm} (f) = \dfrac{S_{nm}(f)}{\sqrt{S_{nn}(f) S_{mm}(f)}},
    \label{eq:StdCoh}
\end{equation}
where $S_{nm} = \mathbf{E} \left\lbrace  \hat p_n(f) \hat p^*_m(f) \right\rbrace$ is the cross-PSD between the $n$-th and $m$-th pressure signals. With the diffuse field model hypothesis, the coherence can be written as a function of only one geometric parameter: $d_{nm} = \| \mathbf{x}_n - \mathbf{x}_m \|$, the distance between microphones $m$ and $n$. This model writes:

\begin{equation}
    \Gamma_{nm}(f) = \text{sinc}\left( \dfrac{2 \pi f d_{nm}}{c_0}\right),
    \label{eq:sinc}
\end{equation}
where $c_0$ is the constant speed of sound. Note that the considered model leads to a real function, even though coherence is a complex function according to its definition.

Other non isotropic plane waves distributions have been studied in \cite{Blake1977}. It is shown that when the plane waves come from a finite solid angle an other analytic model exists whose variations depend on extra geometric parameters: the incoming solid angle aperture and the angle of arrival of the plane waves. More generally, it means that an analytic expression of the coherence function can be derived for other types of scenarios. However it will depend on supplementary geometric parameters, and make the pairwise distance estimation trickier. Another impact on the coherence can come from the microphones directivity \cite{Kuster2008}. They indeed should be omnidirectional so as to render properly the isotropy of the diffuse field.

In \cite{Taghizadeh2014}, Taghizadeh \textit{et al.} makes an exhaustive study of the applicability of the diffuse field model for the geometric calibration of microphone arrays. Although the diffuse field model is not representative of any realistic condition, the study investigates its suitability and limits in various situations. 
The most suitable framework appears to be a large room containing one or some broad band sources and where strong reverberation occurs \textit{i.e.} where direct paths from the sources towards the  microphones are not significant in terms of energy compared with the total acoustic energy \cite{DelGaldo2012}.
Therefore, large rooms with a high level of reverberation are a very convenient environment for array calibration in diffuseness. Nevertheless, the proposed isotropic plane wave model can be met in other situations. In an arbitrary environment with a great number of uncorrelated broadband sources surrounding the array at far field, an acoustic field with the desired properties can also be obtained \cite{Roux2005}. For convenience, the experiments to be described hereafter were performed in a large reverberant hall.

\subsubsection{Coherence estimator}
\label{sec:CoherenceMeasure}

The coherence is computed using the estimator proposed in \cite{Taghizadeh2014}. Let $\hat p^{ST\,b}_m(f)$ denoting the Short Time Fourier Transform (STFT) of the $b$-th data frame of pressure measured by the $m$-th microphone, among $B$ frames. We consider the the following estimator:

\begin{equation}
    \label{eq:estC2}
    \widetilde\Gamma_{nm}(f) =\frac{1}{B}
    \sum_{b=1}^{B} \dfrac{\hat p_n^{ST\,b}(f) \hat p_m ^{ST\,b \:*}(f)}{\vert \hat p_n^{ST\,b}(f) \vert\vert \hat p_m^{ST\,b}(f) \vert} .
\end{equation}
Indeed rather than using the single-frame coherence, Taghizadeh \textit{et al.} show that averaging on several frames improves pairwise distance estimation \cite{Taghizadeh2014}; however it does not eliminate the estimator bias. The STFTs $\hat p_n^{ST\,b}(f)$ are computed using a Blackman taper function which strongly reduces the spectral leakage. Note that our  experimental estimator  $\widetilde {\Gamma}_{nm} (f)$ is intrinsically complex, and presents a non zero imaginary part, while according to the \textit{sinc} fitting model, it should equal zero for all frequencies of interest.  This is due to the imperfect isotropy of the recorded diffuse field. Nevertheless this does not prevent the \textit{sinc} coherence model and the coherence estimate to match well enough to provide reliable estimates of the pairwise distances. Therefore we choose to discard the imaginary part of our estimator, retaining only the real part.

\subsection{Pairwise distance estimation}
\label{sec:EstimationPairwiseDistances}
\subsubsection{Least square minimization}

Considering one pair of microphones, a least square discrepancy minimization process can be implemented between the \textit{sinc} coherence model and the coherence estimate $\widetilde\Gamma_{nm}(f)$. The minimization process providing the estimated pairwise distances $\delta_{nm}$ writes:

\begin{equation}
    \label{eq:leastsq}
    \delta_{nm} = 
    \operatornamewithlimits{argmin}_{d}
    \sum_{f_k} \left| \Re \left[ \widetilde\Gamma_{nm}(f_k) \right] - \text{sinc}\left( \dfrac{2 \pi f_k d}{c_0}  \right) \right|^{2}.
\end{equation} 

This non linear regression can be achieved \textit{e.g.} by a Nelder-Mead search, for all pairwise combinations of microphones. Then, the resulting dataset $\left\lbrace \delta_{nm} \right\rbrace$ contains $M(M-1)/2$ elements.

\subsubsection{Discussion and limitations of the model} 
\label{sec:LimitationsModel}

In this section, the limitations of the \textit{sinc} coherence model are investigated in order to assess which pairwise microphone distances are estimated correctly.

The theoretical $sinc$ function is known to hold 97.9 \% of its energy between the origin and the $4$-th zero. Therefore, the quality of the estimation is evaluated only on this portion. To find criteria on the distances accessible thanks to this model, one requires that the theoretical and experimental coherences fit up to the fourth zero of the $sinc$ function. At this point, the argument of the $sinc$ function is:
\begin{equation}
    \frac{2\pi f d_{nm}}{c_0} = 4 \pi .
    \label{eq:4thZero}
\end{equation}

The pressure field is recorded by finite frames of duration $N/f_s$, where $N$ denotes the number of samples per frame and $f_s$ the sampling rate. Consequently the set of frequency bins ranges from $0$ to $f_s/2$ with a spectral resolution $\Delta f = f_s/N$. Considering the $sinc$ model, these parameters essentially dictate the boundaries of the estimation of $d_{nm}$.

The smallest accessible distance between two microphones is the distance satisfying (\ref{eq:4thZero}), when the frequency is equal to half the sampling rate: 
\begin{equation}
    d_{\text{min}} = \frac{4 c_0}{f_s} .
\end{equation}

At the other bound, for the largest accessible pairwise distance, we impose at least eight bins to discretize the first two oscillations of the $sinc$ function:
\begin{equation}
    d_{\text{max}} = \frac{c_0 N}{4 f_s} .
    \label{eq:dmax}
\end{equation}

Note that these bounds for $\delta_{nm}$ ensue directly from physical causes. The first one is the distance between pairwise microphones: a microphone pair is supposed to listen the same impinging waves coming from all directions. But a time difference of arrival (TDOA) exists due to the distance between the microphones. The maximal value of this TDOA must be negligible \textit{i.e.} small with respect to the frame length:
\begin{equation}
    \label{eq:NecessaryInequality}
    d_{nm} / c_0 \ll \text{N} / f_s .
\end{equation}
This inequality is consistent with the definition (\ref{eq:dmax}) of $d_{\text{max}}$.

Another matter is the validity of the hypothesis on the diffuse nature of the field. In order to assess diffuseness, the Schroeder frequency can be computed. It provides a rough threshold between modal and diffuse behaviors of the ambient acoustic field. It depends on the room volume $V$ and on the reverberation time $T_{60}$ and writes \cite{Schroeder1996}: $f_c = 2000 \sqrt{T_{60} / V}$. It provides another definition of the largest accessible distance with the equation \eqref{eq:4thZero}: 
\begin{equation}
    d_{\text{max}} = \frac{2 c_0 }{f_c} .
\end{equation}

To be consistent with the previous result, the upper bound is now defined by both physical and signal processing considerations:
\begin{equation}
    d_{\text{max}} = \min \left( \frac{c_0 N}{4 f_s};\frac{2 c_0 }{f_c} \right) .
\end{equation}

Note that the bandwidth of microphones has not been considered as a limitation so far. One should make sure of the proper choice of microphones to obtain the measured coherences within the frequency range of interest to estimate the $d_{nm} \in \left[ d_{\text{min}}, d_{\text{max}} \right] $. 

These bounds give an approximate order of magnitude of the limits; yet they show that estimations $\delta_{nm}$ will be awkward for large pairwise distances, and this will be evidenced by experiments in \ref{sec:ExpPairwiseDistance}. Moreover, experimental data exhibit noise and spoil the estimation. This is why a straightforward solution to extract the positions of the microphones from the pairwise distances is not applicable. An intermediary MDS step is required in order to deal with redundant and potentially false data.

\section{Microphones location estimation}
\label{sec:EstimationOfThePositionsOfTheMicrophones}

The unknown microphone positions $\mathbf{x}_1,...,\mathbf{x}_M$ are of dimension $D$. They can be extracted from the set of distance estimates $\left\lbrace \delta_{nm} \right\rbrace$ found previously, whose cardinal is $M(M-1)/2$. However this set lies in a $M$-dimensional space. Therefore the last calibration step consists in finding a projection of the $M$-dimensional input data onto a $D$-dimensional space. This is feasible using MultiDimensional Scaling methods \cite{Cox2000}.

We previously noted that estimating $\delta_{nm}$ between distant microphones of the array is statistically impractical beyond a physical limit $d_{\text{max}}$. In wireless sensor network localization, a similar issue arises: the time of arrival measurements are prone to deteriorating when the pairwise devices are too distant. However some MDS methods are of good interest when the $M$-dimensional input dataset is incomplete \cite{Venna2006,Tenenbaum2000,Leeuw2009}. These MDS algorithms are interesting for practical applications because they rely on the pairwise distances of the closest neighbors. Thus, using the same approach as in sensor localization problems seems appropriate.

However, the error in times of arrival between wireless distant devices usually follows a statistics model. This can be formulated and used efficiently to reduce the calibration error up to the Cramer-Rao bound \cite{Costa2006}. In our case establishing an error model seems inconsistent. Indeed, we will see in the experimental section \ref{sec:Exps} that the wrong data are random outliers, \textit{i.e.} data with a very large error and that they are due to failures in solving the least square minimization problem \eqref{eq:leastsq}. This is problematic since standard MDS methods are known to be strongly sensitive to outliers \cite{Forero2011}, deteriorating the  calibration result.

In the following we propose an enhanced MDS method to overcome the outliers problem. It relies on two key points: first it acts \textit{locally}, secondly, it implements a step to detect and cleanse outliers.

\subsection{The Local and Robust MDS method}

Forero \textit{et al.} \cite{Forero2012} propose a Robust MDS (RMDS) method whose data model is outlier-aware, \textit{i.e.} the outlying estimation errors in $\left\lbrace \delta_{nm} \right\rbrace$ are extended parts of the natural data noise. This brings a threefold expression:

\begin{equation}
    \label{eq:NoiseModel}
    \delta_{nm} = d_{nm} + \epsilon_{nm} + o_{nm}
\end{equation}

where $\left\lbrace \epsilon_{nm} \right\rbrace$ is a subset of zero-mean independent random variables, and $\left\lbrace o_{nm} \right\rbrace$ the set of outlying errors. If the  $\left\lbrace o_{nm} \right\rbrace$ set is sparse \textit{i.e.} if the outliers are few compared with the cardinal of the whole dataset $\left\lbrace \delta_{nm} \right\rbrace$, Forero \textit{et al.} show that it is possible to obtain a consistent estimate $\mathbf{\widetilde X}$ by properly detecting and inhibiting the errors in outliers.

The analysis of an experimental dataset $\left\lbrace \delta_{nm} \right\rbrace$ is detailed in section \ref{sec:ExpPairwiseDistance}. It shows that a number of outliers occur for large distances. Most of them can be rejected by dropping the values higher than a threshold called $\delta_{\text{max}}$. Thus, if some outliers remain in the selected local set of $\left\lbrace \delta_{nm} \right\rbrace$, there should be few, which would guarantee the sparsity hypothesis on $\left\lbrace o_{nm} \right\rbrace$.

For this purpose, we propose a Local and Robust MDS (LRMDS) procedure, based on the RMDS and taking advantage of local considerations. Following the previous rejection step the LRMDS algorithm optimally performs the geometric calibration, by removing the outlying errors to retrieve the \textit{best} input dataset and optimize the estimation of $\mathbf X$. This step of the LRMDS algorithm consists in minimizing the following cost function:

\begin{multline}
    \label{eq:fcoutw}
    f(\mathbf{O}, \mathbf{X}) = \sum_{n<m} w_{nm} [\delta_{nm} - d_{nm}(\mathbf{X}) - o_{nm}]^2\\
    + \nu \sum_{n<m} w_{nm} \vert o_{nm} \vert
\end{multline}

\begin{itemize}
    \item The first term is the quadratic error between $\left\lbrace \delta_{nm} \right\rbrace$, and $\{d_{nm}(\mathbf{X})\}$ (the set of distances derived from the geometric parameter $\mathbf{X}$). Its minimization leads to the best fit of the measured data $\left\lbrace \delta_{nm} \right\rbrace$ into the euclidean space comprising the array. Eliminating $o_{nm}$, this term corresponds to the usual \textit{stress} function. Various MDS methods (the Smacof method \cite{Leeuw2009} for instance) are based on minimizing the \textit{stress} function.
    \item The second term involves the $\ell_1$ norm of $\left\lbrace o_{nm} \right\rbrace$ which constraints the sparsity of the outliers. $\nu$ is the constant regularization parameter of the minimization problem, and controls the sparsity of the solution of $\left\lbrace o_{nm} \right\rbrace$.
\end{itemize}

Function $f$ is similar to the original RMDS cost function, but it is weighted by the $w_{nm}$ coefficients. Our local approach being the total rejection of data higher than the threshold $\delta_{\text{max}}$ whose tuning is described in section \ref{sec:TuningLRMDSParameters}. The weighting strategy follows:

\begin{equation}
    \label{eq:weighting}
    \begin{split}
        w_{nm} &= 1 \text{ if } \delta_{nm} < \delta_{\text{max}} \\
        &= 0 \text{ otherwise}
    \end{split}
\end{equation}

The cost function (\ref{eq:fcoutw}) has two unknowns: the array geometry $\mathbf{X}$ and the sparse matrix of outlying errors $\mathbf{O}$ such that $O[n,m] = o_{nm}$. And finally retrieving the geometry of our microphone array results in solving the following problem:

\begin{equation}
    (\mathbf{\widetilde{O}}, \mathbf{\widetilde{X}}) = 
    \operatornamewithlimits{argmin}_{\mathbf{O} \in \mathbb{S}^M_h , \mathbf{X}}
    f(\mathbf{O}, \mathbf{X})
\end{equation}

where $\mathbb{S}^M_h$ denotes the set of symmetric and hollow matrices of size $M \times M$.

\subsection{Solving the LRMDS problem}

A Majorization-Minimization (MM) approach is capable to overcome the minimization complexity of the cost function \eqref{eq:fcoutw} which is both non-convex and non-differentiable. The MM approach consists in using a surrogate function with better convergence properties for minimization. The algorithm is iterative: at each iteration $k$, a new surrogate function is defined according to the current state of $(\mathbf{O}^{(k)}, \mathbf{X}^{(k)})$, and is minimized, leading to an updated solution $(\mathbf{O}^{(k+1)}, \mathbf{X}^{(k+1)})$. Iteration stops when convergence is verified by a chosen criterion.

One iteration of the algorithm processes in two steps: it updates $\mathbf{\widetilde{O}}$ first, and $\mathbf{\widetilde{X}}$ second. Thanks to the outliers' sparsity hypothesis, the first step can be solved by a standard LASSO method \cite{Buhlmann2011}; indeed the cost function \eqref{eq:fcoutw} is equal to a sum of $M(M-1)/2$ scalar LASSO sub-problems on the $o_{nm}$ variables. Updating the latter is possible by using the \textit{soft-thresholding} operator such as:

\begin{equation}
    \label{eq:Otp1}
    o_{nm}^{(k+1)} = S_{\nu}
    \left(
    \delta_{nm} - d_{nm}(\mathbf{X}^{(k)})
    \right)
\end{equation}
with:
\begin{equation}
    \label{eq:SoftTh}
    S_{\nu} (u) = \text{sign}(u) \,
    \text{max} \left\lbrace \vert u \vert - \nu/2 , 0 \right\rbrace 
\end{equation}
Note that the effect of the regularization parameter $\nu$ from the cost function \eqref{eq:fcoutw} applies here, in solving the LASSO problem on $o_{nm}$ variables.
Applying this update for all pairs $(n,m)$ gives the updated matrix $\mathbf{O}^{(k+1)}$.

The second step updating $\mathbf{\widetilde{X}}$, demonstrated in \cite{Forero2012}, derives from:

\begin{equation}
    \label{eq:Xtp1}
    \mathbf{X}^{(k+1)} = \mathbf{X}^{(k)} 
    \mathbf{L_1} (\mathbf{O}^{(k+1))}, \mathbf{X}^{(k)} ) 
    \mathbf{L^{\dagger}}
\end{equation}
Our contribution in the process relies in introducing the weighting coefficient $w_{nm}$. In the present case of the weighted cost function, $\mathbf{L_1}(\mathbf{O}, \mathbf{X})$ writes:

\begin{equation}
    \label{eq:L1}
    \mathbf{L_1}(\mathbf{O}, \mathbf{X}) = \text{diag} (
    \mathbf{A_1}(\mathbf{O}, \mathbf{X})\mathbf{1}_M
    )
    - \mathbf{A_1}(\mathbf{O}, \mathbf{X})
\end{equation}
with, $\mathbf{1}_M$ the column vector of $M$ ones, $\text{diag}(v)$ the square diagonal matrix containing the elements of $v$ in the diagonal, and: 
\begin{equation}
    \begin{split}
        [\mathbf{A_1}(\mathbf{O}, \mathbf{X})]_{nm} 
        &= 
        w_{nm} \dfrac{\delta_{nm} - o_{nm}}{d_{nm}(\mathbf{X})} \quad \text{if $ \delta_{nm} > o_{nm}$} \\
        &\hspace{3.25cm} \text{and $d_{nm}(\mathbf{X}) > 0$} \\
        & = 0 \hspace{2.6cm} \text{otherwise}
    \end{split}
\end{equation}

Jointly, $\mathbf{L^{\dagger}}$ is the pseudo-inverse matrix of $\mathbf{L}$ such that:
\begin{equation}
    \label{eq:L}
    \mathbf{L} = \sum_{n<m} 
    w_{nm} (\mathbf{e}_n - \mathbf{e}_m)^T (\mathbf{e}_n - \mathbf{e}_m)
\end{equation}
with $\mathbf{e}_n$ the row vector of $M$ elements, defined as $[\mathbf{e}_n]_m = 1$ if $m=n$ , and 0 otherwise.

Equations \eqref{eq:Otp1}-\eqref{eq:L} describe the computations of one iteration of the LRMDS. The algorithm iterates as long as convergence is not guaranteed; for that the following ratio is computed \cite{Forero2012}: $ \| \mathbf{X}^{(k+1)} - \mathbf{X}^{(k)} \|_{F} 
/
\| \mathbf{X}^{(k)} \|_{F}$, with $\|.\|_{F}$ the Frobenius norm. If it is lower than a certain threshold \textit{e.g.} $10^{-6}$, then the algorithm stops. The last computed couple $(\mathbf{O}^{(k)}, \mathbf{X}^{(k)})$ gives the final estimate $(\mathbf{\widetilde O}, \mathbf{\widetilde X})$.

\bigskip

To conclude, the overall proposed LRMDS method can be summarized by the following key points:

\begin{enumerate}
    \item From the $\left\lbrace \delta_{nm} \right\rbrace$ set, only the local values are selected, so that $\delta_{nm} < \delta_{\text{max}}$. It leads to the subset $\left\lbrace \delta_{nm} \right\rbrace_{< \delta_{\text{max}}}$.
    \item From $\left\lbrace \delta_{nm} \right\rbrace_{< \delta_{\text{max}}}$, the LRMDS algorithm solves the calibration problem, and detects the existing outliers in $\left\lbrace \delta_{nm} \right\rbrace_{< \delta_{\text{max}}}$. The corresponding outlying errors are collected in the $\left\lbrace o_{nm} \right\rbrace$ set.
    \item The final estimate $\mathbf{\widetilde X}$ results from the best fit, into the euclidean space comprising the array, of the $\left\lbrace \delta_{nm} \right\rbrace_{< \delta_{\text{max}}}$ subsets, having cleansed the large errors in $\left\lbrace o_{nm} \right\rbrace$.
\end{enumerate}

\section{Experiments}
\label{sec:Exps}

In order to validate the overall process of figure \ref{fig:CalibChart}, we perform a two step investigation. First a global analysis of a raw pairwise distance set is achieved, in order to investigate the limitations of the fitting process of experimental coherences as stated in section \ref{sec:LimitationsModel}. Second, the implementation of the LRMDS algorithm is described for a real case, and its performances are studied considering two different microphone arrays. 

\subsection{Experimental setup}

\subsubsection{Measuring the soundfield}

The acoustic acquisition system enables to set up an array containing up to 128 digital microphones. The latter are made up by implementing the ADMP441 component from Analog Device which is based on a MEMS architecture, and straightforwardly delivers digital samples of the measured acoustic pressure in output. The acquisition system is fully digital which is convenient for the robust data transfer over long distances, and thus for the spatial extent of the array. The MEMS components development are mainly meant for massive integration in mobile phones. However their performances are sufficient for standard microphone array applications. An acoustic characterization of these sensors, as well as the hardware architecture of the apparatus, are exhaustively described in \cite{Vanwynsberghe2015}. 

In our experiments, the sampling frequency of the acoustic signals is set to $f_s=50$~kHz allowing the coverage of the audible domain and the working bandwidth of the microphones as well.

All the experiments are achieved in a large hall ($10 \times 8 \times 20$~m$^3$); based on the measured reverberation time, its Schroeder frequency is $f_c = 77$~Hz. The soundfield comes from different natural sources (speeches, machines, and so on ...). The coherences are computed using frames of $N = 2048$ samples at $f_s=50$~kHz. According to section \ref{sec:LimitationsModel}, this choice theoretically limits the pairwise distances estimation to the interval $ d_{nm} \in \left[ 2.7 \text{~cm} ; 3.5 \text{~m} \right] $. The total acquisition duration is set to 1 minute; with the chosen frame length $N$, the coherences are computed by averaging $B = 1464$ frames. A longer acquisition time does not improve the estimated dataset $\left\{ \delta_{nm} \right\}$ (and does not degrade the result either).

\subsubsection{Antenna geometries}

\begin{figure}
    \centering
    \begin{subfigure}[b]{0.4\textwidth}
        \centering
        \includegraphics[width=\textwidth]{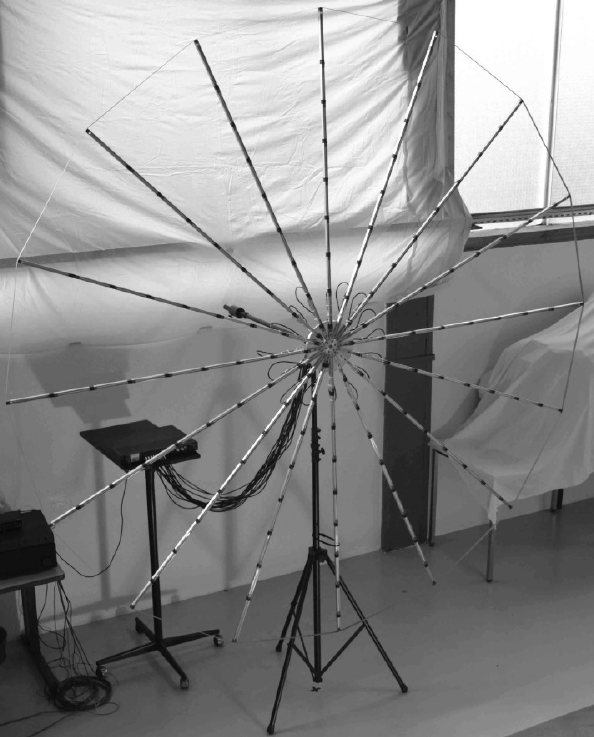}
        \caption{Array 1}
        \label{fig:Arrays-Circular}
    \end{subfigure}
    \begin{subfigure}[b]{0.4\textwidth}
        \includegraphics[width=0.97\textwidth]{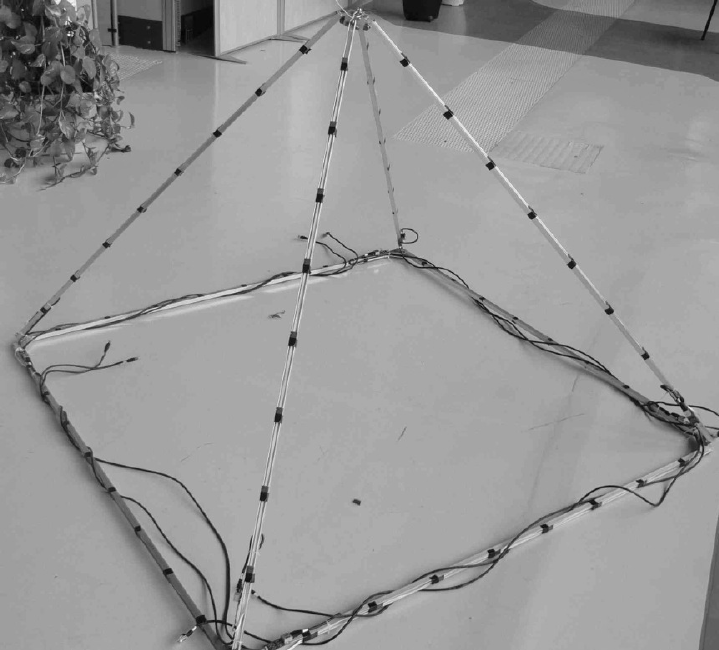}
        \caption{Array 2}
        \label{fig:Arrays-Pyramid}
    \end{subfigure}
    \caption{Arrays for the validation experiments. In \ref{fig:Arrays-Circular}: circular array, in \ref{fig:Arrays-Pyramid}: pyramid array.}
    \label{fig:Arrays}
\end{figure}

The overall performances of the method are investigated through two scenarios. They keep an identical experimental background, but implement two different arrays:

\begin{itemize}
    \item Array 1 is a circular array of 128 MEMS microphones (Fig. \ref{fig:Arrays}), typically used for acoustic imaging. It is made up of 16 bars of 8 microphones. Each bar is a radius of a disk, $2.5$~m wide in diameter, as shown in figure \ref{fig:Arrays}. These dimensions imply relatively large pairwise distances up to $2.5$~m. This 2D configuration is therefore very interesting to investigate the efficiency of both the  \textit{local} and the \textit{outlier-aware} steps of the LRMDS method.
    \item Array 2 is a square pyramid (Fig. \ref{fig:Arrays}). Each edge is  $1.5$~m long and holds 8 microphones regularly set, resulting in a 3D array of 64 microphones. This geometry ensures an accurate knowledge of the real position of the microphones, down to the millimeter.
\end{itemize}

\subsection{Experimental estimation of the pairwise distances}
\label{sec:ExpPairwiseDistance}

In this section we describe the experimental process that leads to the dataset of estimated pairwise microphone distances. This dataset is to be the input to the final calibration step providing the geometry. Its reliability must therefore be assessed. The limitations of this step are compared with the statements in section \ref{sec:LimitationsModel}.

\begin{figure*}[t]
    \begin{center}
        \includegraphics[width=0.9\textwidth]{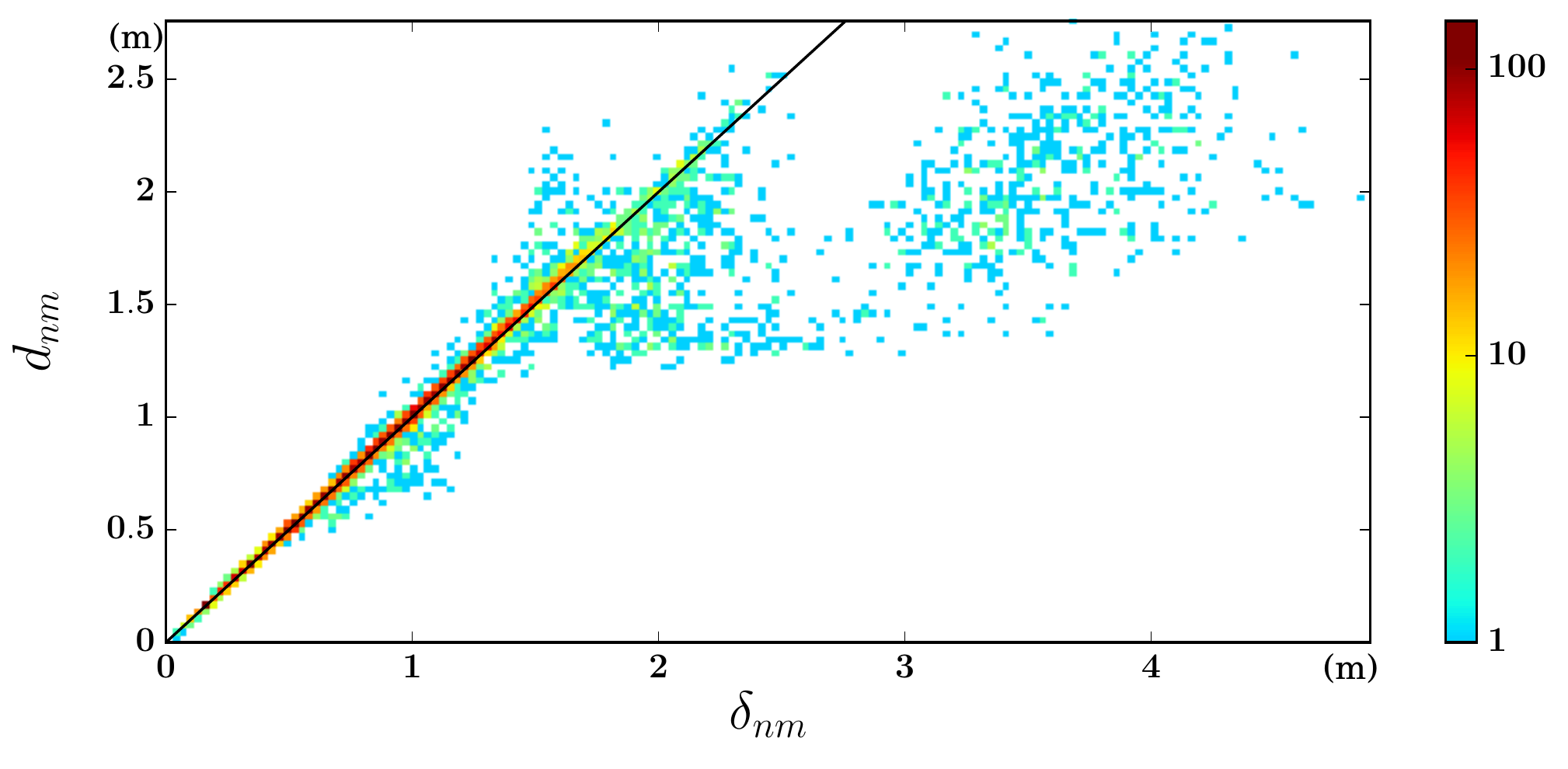}
    \end{center}
    \caption{2D-histogram of $d_{nm}$ vs $\delta_{nm}$ for array 1; the absolute error on the estimation $| \delta_{nm} - d_{nm} |$ is the distance of bins to the line $d_{nm}=\delta_{nm}$. $8128$ pairwise distance estimates. Bin size: $3$~cm $\times$ $3$~cm. (color online)}
    \label{fig:hist2D}
\end{figure*}

The estimation error of the $\delta_{nm}$ derived from the first scenario (circular array) is presented in figure \ref{fig:hist2D}. It plots the $\delta_{nm}$ versus the $d_{nm}$ of the set of $8128$ distances, by a 2D-histogram. The error $| \delta_{nm} - d_{nm} |$ is then illustrated by the distance between the histogram bins and the line $d_{nm}=\delta_{nm}$. The results evidence three points:
\begin{itemize}
    \item The bins with the highest counts exist for $\vert \delta_{nm} - d_{nm} \vert < 2$~cm. This interval counts $77.9 \%$ of the set $\left\{ \delta_{nm} \right\}$. It shows that most distances are correctly estimated. But the other estimated distances have absolute errors strongly and randomly scattered. These can be considered outliers.
    \item The presence of outliers correlates with the value of $\delta_{nm}$, and increases together with $\delta_{nm}$. Indeed, a sharp increase of outliers is observable at $\delta_{nm}=1.5$~m. 
    \item The probability of outliers strongly depends on the value of $d_{nm}$: it increases together with $d_{nm}$. Indeed, distances such that $\vert \delta_{nm} - d_{nm} \vert > 2$~cm represents $22.1 \%$ of the total data; for $d_{nm}<1$~m, it represents $8.4 \%$.
\end{itemize}

The second observation is in agreement with the statements of section \ref{sec:EstimationPairwiseDistances}: estimation of high pairwise distances is critical. It is actually translated into failures of the coherence fitting step, leading to high errors in distance estimation. Consequently estimation of large distances is poorly reliable because of the high probability of outliers. Moreover, a few outliers still remain for smaller distances. This confirms the need, in the final calibration step, to set up a strategy capable of enhancing the robustness. Finally the third observation validates the interest of using a local method in LRMDS: truncating large values of $\delta_{nm}$ enables to remove the largest part of outliers. 

\subsection{Tuning the LRMDS parameters}
\label{sec:TuningLRMDSParameters}

\begin{figure*}[t]
    \begin{center}
        \includegraphics[width=0.9\textwidth]{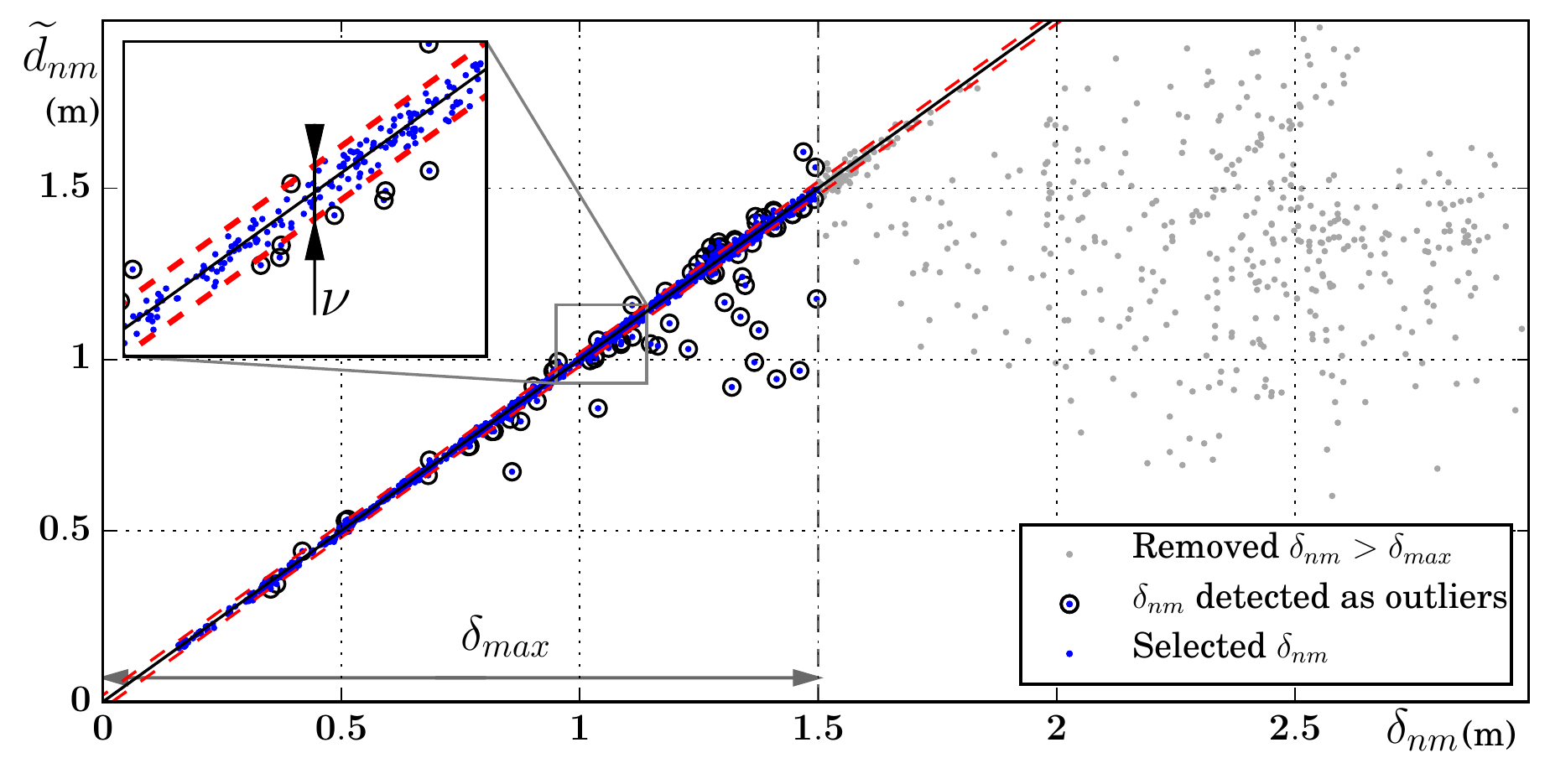}
    \end{center}
    \caption{Geometric calibration of pyramid array 2: Shepard diagram of the LRMDS. (color online)}
    \label{fig:CalibPyramide-Shepard}
\end{figure*}

In this section the LRMDS method is applied completely considering only the pyramid array (array 2). It can be described by functional features: the input data is $\left\lbrace \delta_{nm} \right\rbrace$, the output data is the estimated $\mathbf{\widetilde X}$. Besides, three parameters $D$, $\delta_{\text{max}}$ and $\nu$ drive the algorithm. The first one, generally known \textit{a priori}, is the space dimension of the array ($D=1$, $2$ or $3$). The geometrical interpretation of the second and third parameters is possible reasoning from figure \ref{fig:CalibPyramide-Shepard}; this Shepard diagram \cite{Borg2005} is a scatter plot of the input distances $\delta_{nm}$ versus the distances output from the estimated geometry $\mathbf{\widetilde X}$: $\widetilde d_{nm} = \| \mathbf{\widetilde x}_n - \mathbf{\widetilde x}_m \|$. Therefore this diagram is observable in a blind calibration process, and is rich in information to assess the result of the LRMDS. It shows the relationship between the MDS input data and its final projection found by the algorithm. Its purpose is to evaluate how the MDS estimated geometry matches the input distance set: in an ideal calibration process with faultless input data, all the points would be located on the line $\widetilde d_{nm} = \delta_{nm}$. With real data though, most of the scatter points are expected to be close to this line when the geometric calibration process succeeds. The points located far from this line reflect a distortion between the input and output data. This can be due to two reasons:

\begin{itemize}
    \item the $\widetilde d_{nm}$ output by the MDS process is uncorrelated with the input data. 
    
    \item the measured $\delta_{nm}$ contains a large error.
\end{itemize}

Here, the latter reason is relevant to understand the outlier detection in the proposed LRMDS strategy: the algorithm detects the $\delta_{nm}$ located far from the $\widetilde d_{nm} = \delta_{nm}$ line. In figure \ref{fig:CalibPyramide-Shepard}, the points detected as outliers and rejected for calibration are circled. The dataset retained for calibration is located between the lines $\widetilde d_{nm} - \nu / 2 < \delta_{nm} < \widetilde d_{nm} + \nu / 2$. This shows the effect of the soft-thresholding operator in the algorithm, given in equation \eqref{eq:SoftTh}. Then the regularization parameter $\nu$, in meters, sets the limit of the acceptable error in the estimated pairwise distances.

\begin{figure}
    \begin{center}
        \includegraphics[width=0.9\linewidth]{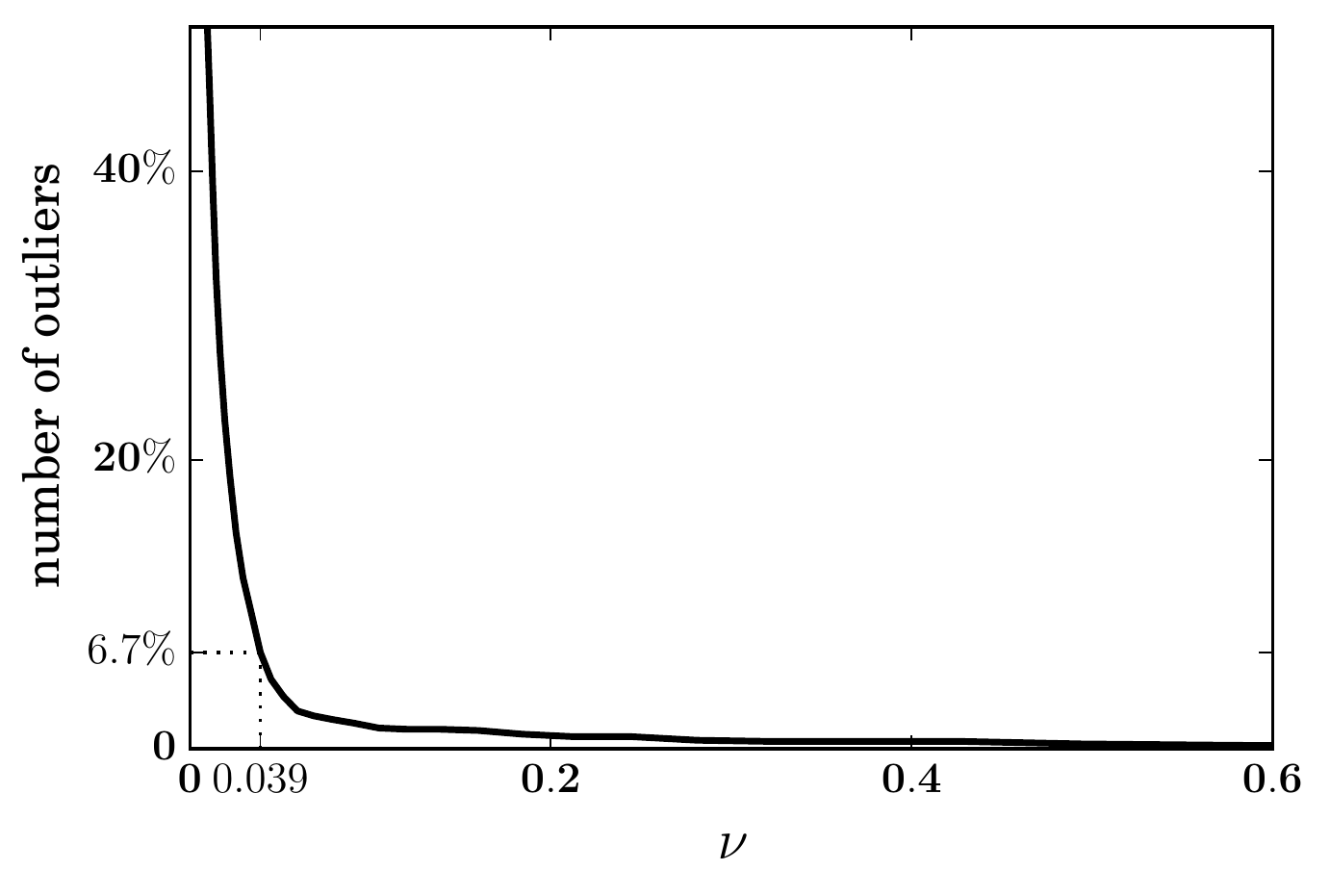}
    \end{center}
    \caption{Geometric calibration of pyramid array 2: tuning the regularization parameter $\nu$ in LRMDS, for $\delta_{\text{max}} = 1$~m.}
    \label{fig:CalibPyramide-LambdaTune}
\end{figure}

The objective choice of $\nu$ is achievable by processing multiple LRMDS with different values. This results in the plot of figure \ref{fig:CalibPyramide-LambdaTune}, showing the cardinal of $\left\lbrace o_{nm} \right\rbrace$ \textit{i.e.} the number of detected outliers, function of $\nu$. Reducing $\nu$ loosens the constraint and increases the outlier detection sensitivity. So we observe that the cardinal of $\left\lbrace o_{nm} \right\rbrace$ increases when $\nu$ decreases. The curve appears to show an L-shape, on which two distinct parts can be interpreted.
On the right of the knee, $\nu$ has relatively high values, which entails a strong constraint on the outliers sparsity. Consequently only a few samples are detected as outliers. 
On the left of the knee, the number of detected outliers strongly increases when $\nu$ decreases: according to equation \eqref{eq:NoiseModel}, $\delta_{nm}$ also contains an error $\epsilon_{nm}$ which can lead to the false detection of an outlier if $\nu$ is too small. Therefore, the \textit{correct} choice for $\nu$ should lie between these two areas (\textit{i.e.} at the knee of the L-curve), where the maximum of actual outliers can be detected.

Lastly, the local selection process of LRMDS rejects all the $\delta_{nm}$ greater than $\delta_{\text{max}}$ (gray dots in figure \ref{fig:CalibPyramide-Shepard})  which are not involved in the minimization process of function \ref{eq:fcoutw}. Unlike the selection of the $\nu$ parameter, no adequate parametric study enables an optimum choice of $\delta_{\text{max}}$. But it should be judicious to take into account that:
\begin{itemize}
    \item the sparsity of outliers in the local subset $\left\lbrace \delta_{nm} \right\rbrace_{< \delta_{\text{max}}}$ must be guaranteed for the success of the LRMDS. In practice, too much outliers results in an algorithm convergence failure. On the first hand, a necessary condition was evidenced: pairwise distances cannot be estimated over a physical limit $ d_{\text{max}}=3.5$~m. On the other hand, in experimental section \ref{sec:ExpPairwiseDistance} we evidenced that the presence of outliers in $\left\lbrace \delta_{nm} \right\rbrace_{< \delta_{\text{max}}}$ is significantly reduced compared with $\left\lbrace \delta_{nm} \right\rbrace$, by choosing $\delta_{\text{max}} \approx 1.5$~m.
    
    \item However reducing $\delta_{\text{max}}$ means, by the equation \eqref{eq:weighting}, weighting more terms to zero in the minimized cost function $f$. This degrades the convexity property, and results in a minimization process more sensitive to local minima. Then the algorithm can converge to an inconsistent solution $\mathbf{\widetilde{X}}$.
\end{itemize}

Finally, the proper value for $\delta_{\text{max}}$ is the highest for which the algorithm succeeds to converge. 
The parameters chosen in the case of the two arrays are listed in table \ref{tab:parameters}, together with the resulting size of the subset $\left\{\delta_{nm}\right\}_{< \delta_{\text{max}}}$ and the number of outliers which is detected there. It shows that the experimental geometric calibrations succeeds with a range capping of large distances beyond $\delta_{\text{max}} = 1$~m or $1.5$~m.

\begin{table}[h]
    \centering
    \caption{Parameters for the two geometrical calibration experiments. The resulting size of the local subset $\left\{\delta_{nm}\right\}_{< \delta_{\text{max}}}$ and the number of detected outliers are given in \%.}
    \begin{tabular}{c|c c c c|c|c}
        \hline 
        Array 	& $M$	& $D$	& $\delta_{\text{max}}$ (m) & $\nu$ (m) & $\frac{\# \left\{\delta_{nm}\right\}_{< \delta_{\text{max}}}}{\# \left\{\delta_{nm}\right\}}$ 	&  $\frac{\#\left\{o_{nm}\right\}}{\#\left\{\delta_{nm}\right\}_{< \delta_{\text{max}}}}$	\\ [2ex]
        \hline 
        1 		& $128$	& $2$	& $1$	& $0.043$	& $45.1\%$	& $6.6\%$	\\ 
        \hline
        2 		& $64$ 	& $3$	& $1$ 	& $0.039$ 	& $54.8\%$	& $6.7\%$	\\ 
        \hline 
    \end{tabular}
    \label{tab:parameters}
\end{table}

\subsection{Results and discussion}
\label{sec:ResultsDiscussion}

In order to quantify the efficiency of the calibration procedure, the microphone positioning error is calculated. The proposed method is compared with the Taghizadeh \textit{et al.} one \cite{Taghizadeh2014} by the results, to evidence the differences of both approaches. Afterwards the impact of the geometry error is evaluated in the frame of a simulated acoustic imaging scenario.

\subsubsection{Geometry estimation performance}

\begin{figure*}
    \centering
    \begin{subfigure}[b]{0.46\textwidth}
        \centering
        \includegraphics[width=\linewidth]{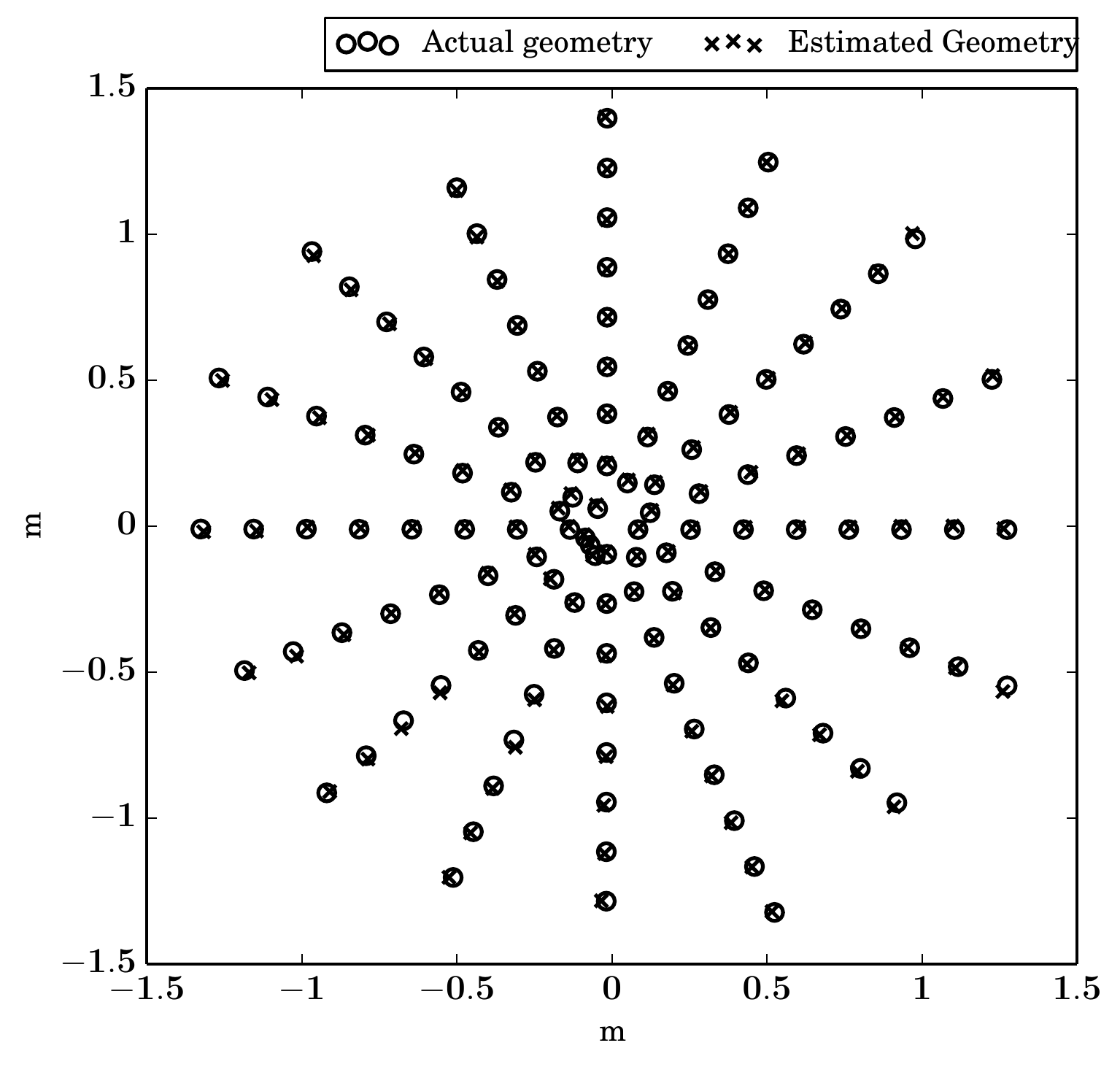}
        \caption{Array 1}
        \label{fig:Calib-Procrustes-Soleil}
    \end{subfigure}
    \begin{subfigure}[b]{0.46\textwidth}
        \centering
        \includegraphics[width=\linewidth]{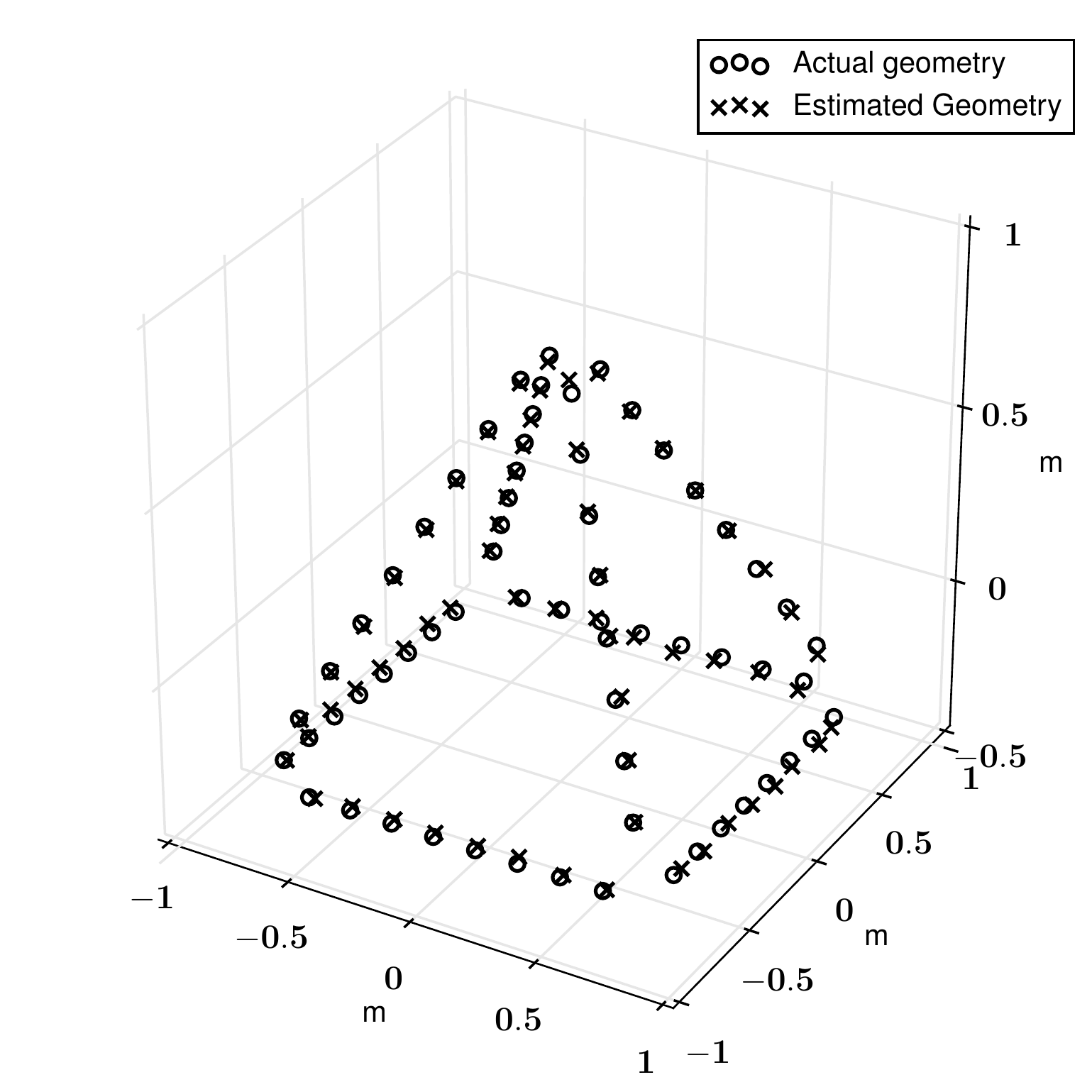}
        \caption{Array 2}
        \label{fig:Calib-Procrustes-Pyramid}
    \end{subfigure}
    \caption{Geometric calibration results, with the proposed method: comparison of the actual and estimated geometries.}
    \label{fig:Calib-Procrustes}
\end{figure*}

The estimated geometry is aligned onto the real one by means of a Procrustes analysis \cite{Gower2010}. This geometrical technique consists of removing the existing translations, rotations, and scale discrepancies between $\mathbf{\widetilde{X}}$ and $\mathbf{X}$ \cite{gower2004procrustes}. Translations and rotations inevitably exist since the LRMDS algorithm results in an estimated geometry $\mathbf{\widetilde{X}}$ that depends only on the relative positions of the microphones. Besides, a scale ratio would \textit{de facto} issue from an error on the value of the speed of sound $c_0$ used to estimate the  $\delta_{nm}$, even though the impact on the positioning error is negligible. Following the Procrustes transformation, figure \ref{fig:Calib-Procrustes} exhibits the final result for the two arrays.

The error per microphone is chosen to be the euclidean distance between the actual and the estimated microphone position: $\varepsilon_n = \lVert \mathbf{\widetilde x}_n - \mathbf{x}_n \rVert$. The results for the two configurations are given in table \ref{tab:error} in term of global, minimal, and maximal errors, as well as the standard deviations. 

\begin{table}[h]
    \centering
    \caption{Overall geometry error for the two arrays. Average, minimum, maximum, and standard deviation in cm.}
    \begin{tabular}{c c c c c}
        \hline Array 	& $\overline \varepsilon_n$	& $\min(\varepsilon_n)$ & $\max(\varepsilon_n)$ & $\text{std}(\varepsilon_n)$	 \\ 
        \hline 1 		& $0.94$ 					& $0.08$		 		& $2.8$ 				& $0.50$	\\ 
        2 		& $2.2$ 					& $0.85$ 				& $4.6$ 				& $0.87$  	\\ 
        \hline 
    \end{tabular}
    \label{tab:error}
\end{table}

For the same experimental background, the calibration process performs better on the circular array than on the pyramidal array. Although the former is larger and wider than the latter, its microphones lie on a 2D shape, and their distribution is denser: the microphones are more surrounded by close neighbors. Moreover, the microphones directivity could explain the difference of results; let us remind that the microphones should be omnidirectional so as to properly render the isotropy of the diffuse field. In the case of the used MEMS microphones however, the directivity is guaranteed uniform on the $180 \degree$ front angle \cite{Vanwynsberghe2015}, while the rear directivity is unknown. However, the microphones of array 1 are all oriented in the same direction, unlike those of array 2 (cf. fig. \ref{fig:Arrays}). Then the impact of the potential directivity would be higher for array 2.

\begin{figure*}[t]
    \centering
    \begin{subfigure}[b]{0.46\textwidth}
        \centering
        \includegraphics[width=\textwidth]{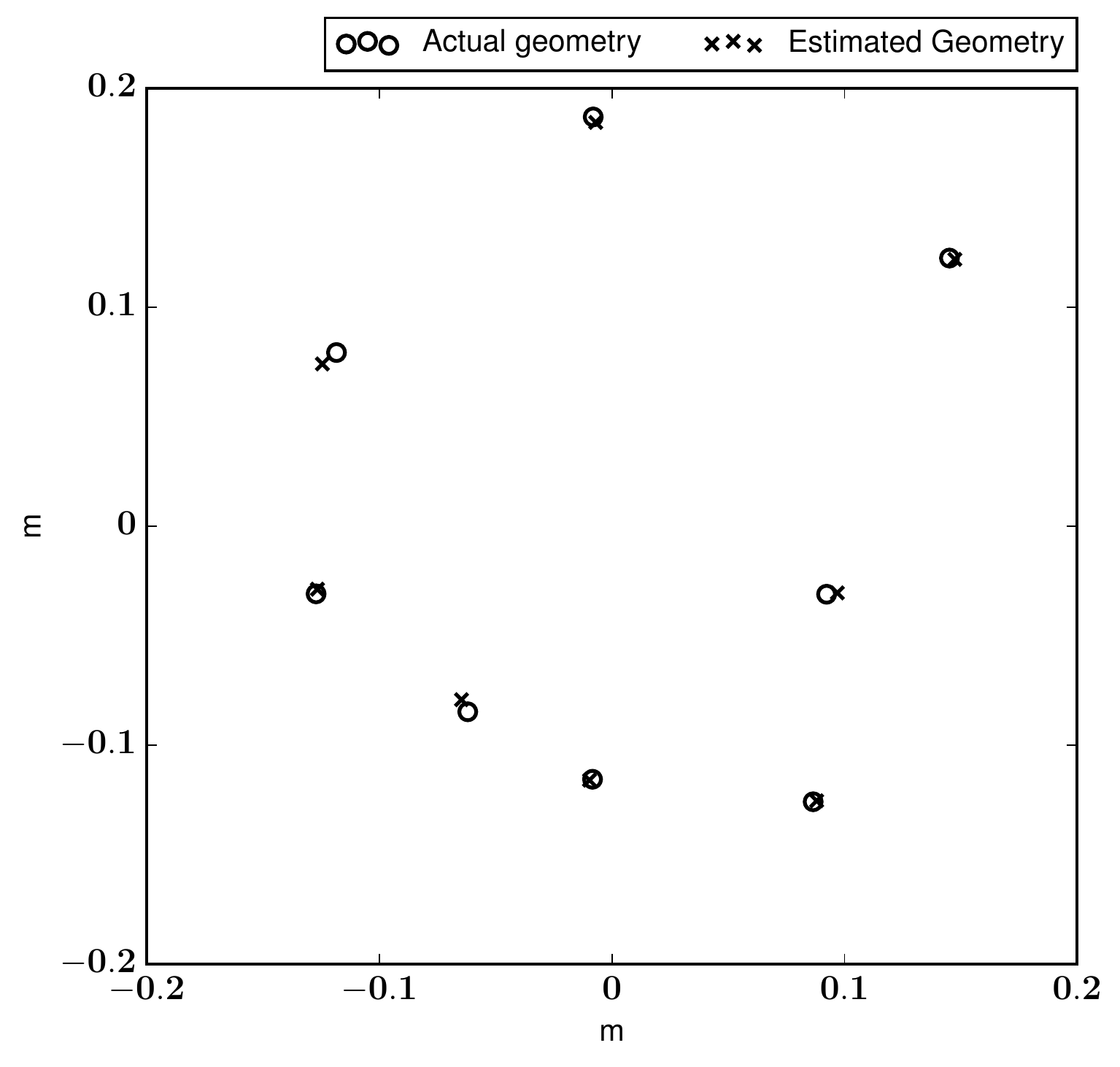}
        \caption{8 microphones at the center of Array 2}
        \label{fig:Calib-Taghizadeh-Small}
    \end{subfigure}
    \begin{subfigure}[b]{0.46\textwidth}
        \centering
        \includegraphics[width=\textwidth]{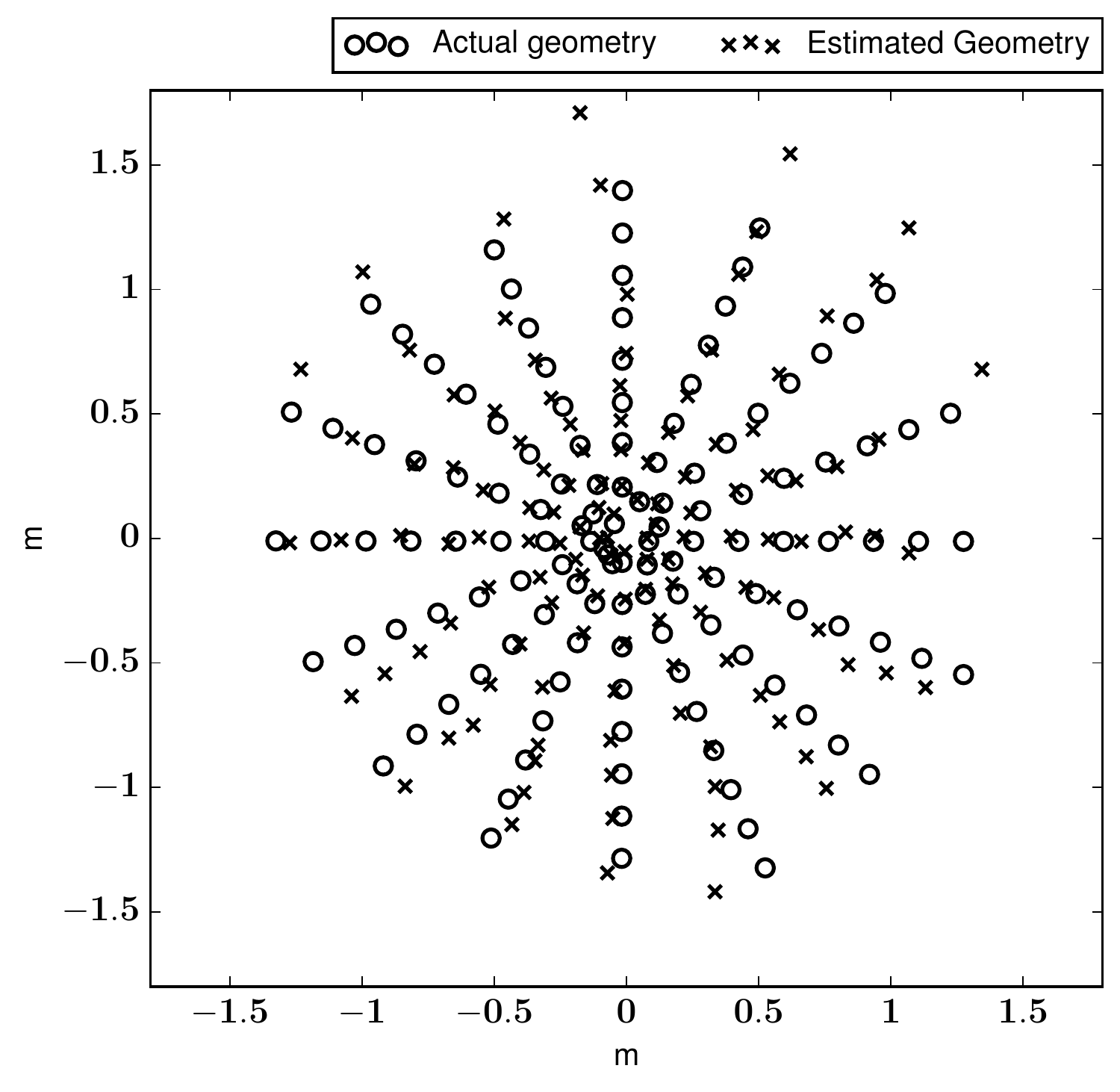}
        \caption{Array 2}
        \label{fig:Calib-Procrustes-Taghizadeh-Soleil}
    \end{subfigure}
    \caption{Geometric calibration results, with 2D-histogram method \cite{Taghizadeh2014}: comparison of the actual and estimated geometries.}
    \label{fig:Calib-Taghizadeh}
\end{figure*}

\subsubsection{Method comparison}
The state of the art methods \cite{McCowan2008,Taghizadeh2014} and the proposed one differ in the strategy to handle the outliers. Taghizadeh \textit{et al.} propose to build the pairwise distances set $\left\lbrace \delta_{nm} \right\rbrace$, by means of a clustering method, prior to processing the MDS. 
For each microphone pair, coherence computation and fitting are achieved on successive frames of the recording ($\approx 400$ times). The clustering relies on a 2d-histogram of the pairwise distances versus the final fitting minimization error (eq. \eqref{eq:leastsq}). The bin with the highest count is the selected pairwise distance value added to $\left\lbrace \delta_{nm} \right\rbrace$. The final geometry estimation is achieved on this set by classic MDS. The experimental work shows good results for a $20$ cm circular array of $9$ microphones \cite{Taghizadeh2014}, but does not exhibit the applicability to large arrays. In the present study, this approach is implemented to assess the geometric calibration performance for two arrays: 
\begin{itemize}
    \item a small array with $M=8$ elements within a disc whose diameter is $15$ cm. 
    \item a large array: the Array 2 (fig. \ref{fig:Arrays-Circular}) used in the previous experiment ($M=128$, diameter $3$ m).
\end{itemize}
Note that the small array consists of the 8 central microphones of the large one. For the two arrays, the 2d-histograms are built by computing and fitting $452$ coherences for each microphone pair. The computational cost for this method is high, growing as $ M^2 $. The estimated geometries are plotted in figure \ref{fig:Calib-Taghizadeh}. The $8$-microphones array is recovered with an average error $\overline \varepsilon_n = 1.1$ mm, which confirms the efficiency of the approach for small arrays. However, the geometry estimation for array 2 is obtained with an average error of $\overline \varepsilon_n = 18.4$ cm, which is $20$ times larger than the one obtained with the Local RMDS proposed in this paper. This is due to the inefficiency of the 2d-histogram clustering which does not discard all the outliers when dealing with large pairwise distances. Yet classic MDS is very sensitive to outliers resulting in a larger estimation error. In comparison, the proposed method proves to need far less computation and to be robust to outliers. Indeed only one coherence computation and fitting per microphone pair is needed to get $\left\lbrace \delta_{nm} \right\rbrace$. Instead of a prior clustering step, it explicitly identifies and efficiently removes the outliers from $\left\lbrace \delta_{nm} \right\rbrace$ and finally provides good estimated geometries even for large arrays.

\subsubsection{Applicability to acoustical imaging}

Whether these errors are acceptable or not depends on the application of the array. We choose to evaluate their impact in the frame of a simulated acoustic imaging scenario. Previous simulation studies have proven that the classic beamformer degrades if a zero mean Gaussian-distributed random error is added on the microphone positions \cite{Sachar2005}. This degradation also increases towards higher frequencies. For our assessment step, we propose to simulate a beamformer using the estimated geometries, in comparison with the actual ones. A source is located $5$~m from the centroid of both antennas. Its signal, whose spectrum is uniform over the band of interest, is measured by the microphones located according to geometry $\mathbf{X}$. Then the beamforming in geometrical Near-Field is computed in the frequency domain, by using (a) the actual geometry $\mathbf{X}$, (b) the geometry provided by the calibration process $\mathbf{\widetilde X}$.

\begin{figure*}
    \centering
    \begin{subfigure}[b]{0.4\textwidth}
        \includegraphics[width=\textwidth]{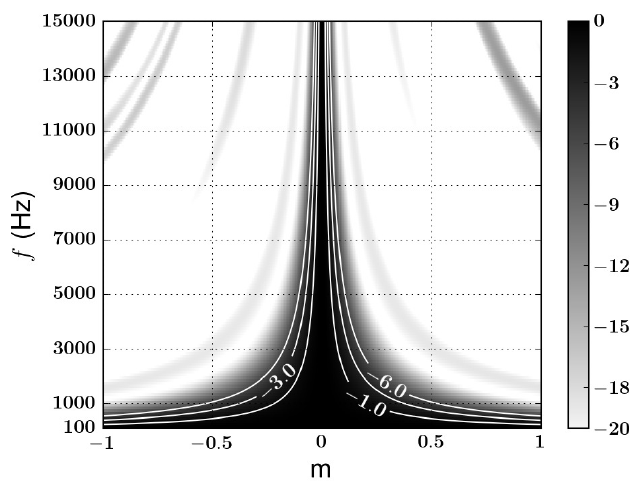}
        \caption{Beamforming reconstructed with the real antenna geometry $\mathbf{X}$, in dB.}
        \label{fig:BeamPattern-Sol-Th}
    \end{subfigure}
    ~ 
    \begin{subfigure}[b]{0.4\textwidth}
        \includegraphics[width=\textwidth]{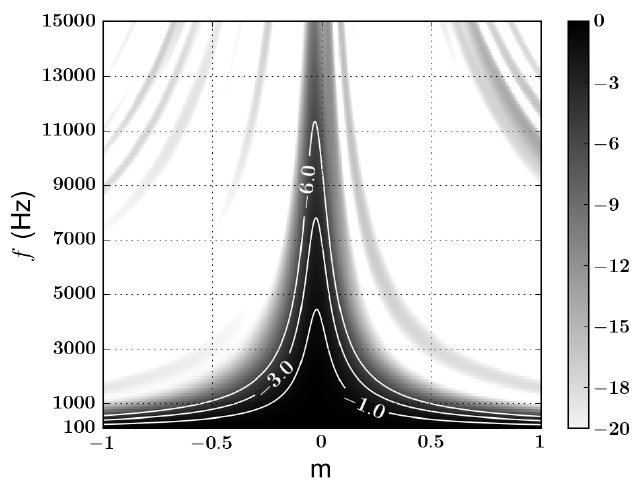}
        \caption{Beamforming reconstructed with the calibrated antenna geometry $\mathbf{\widetilde X}$, in dB.}
        \label{fig:BeamPattern-Sol-Cal}
    \end{subfigure}
    
    \begin{subfigure}[b]{0.4\textwidth}
        \includegraphics[width=\textwidth]{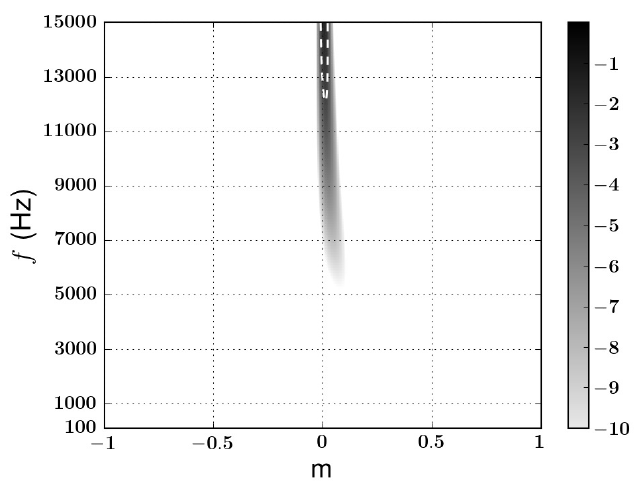}
        \caption{Error on beamforming between \ref{fig:BeamPattern-Sol-Th} and \ref{fig:BeamPattern-Sol-Cal}, in dB. $-3$~dB isocontour in white dashed line.}
        \label{fig:BeamPattern-Sol-Err}
    \end{subfigure}
    \begin{subfigure}[b]{0.4\textwidth}
        \includegraphics[width=6.5cm]{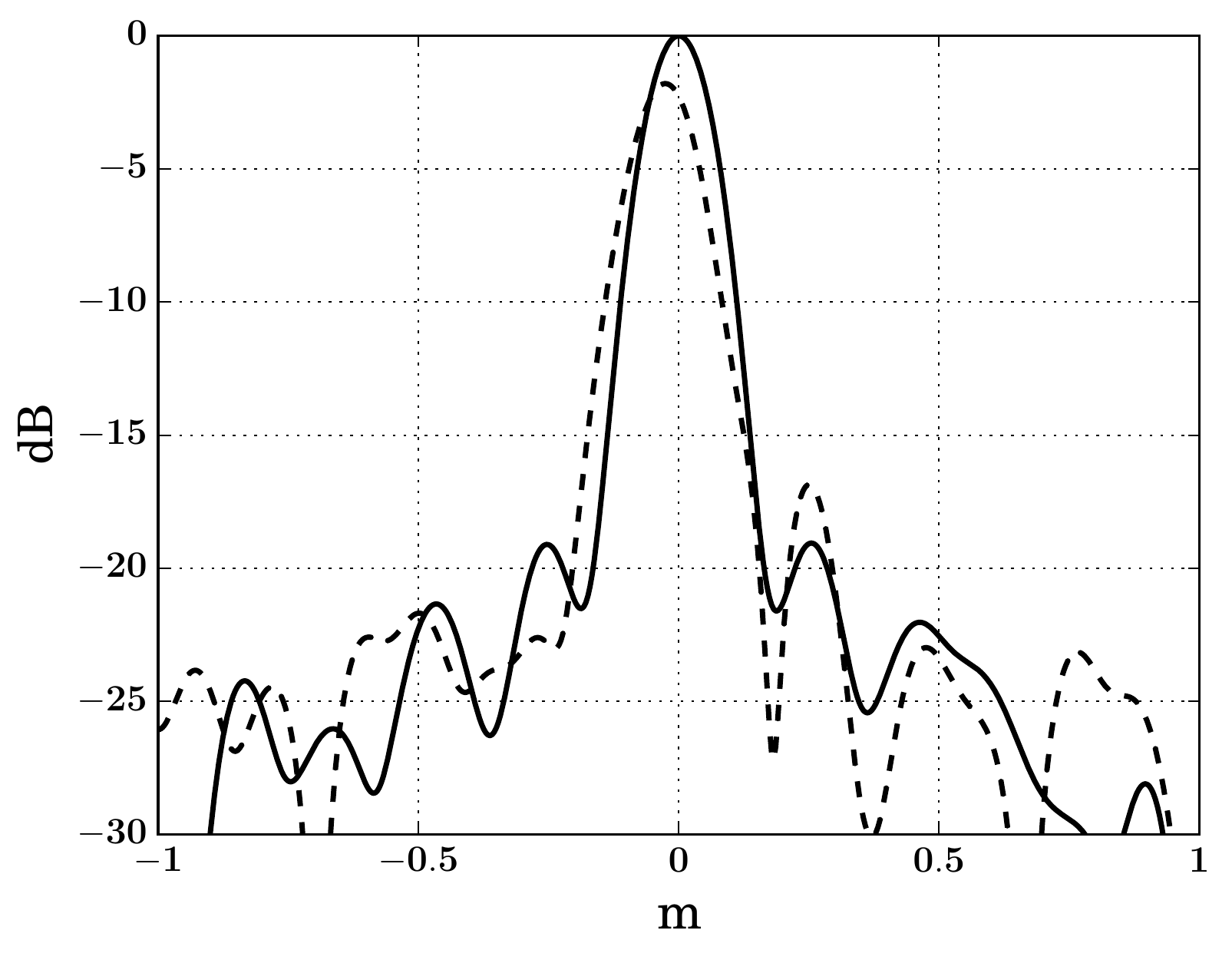}
        \caption{Slice of figures \ref{fig:BeamPattern-Sol-Th} and \ref{fig:BeamPattern-Sol-Cal} for $f=6$~kHz (actual geometry in solid line, calibrated geometry in dashed line).}
        \label{fig:BeamPattern-Sol-6kHz}
    \end{subfigure}
    \caption{Beamforming in geometrical Near-Field with array 1, with a broadband point source at $5$~m from the array centroid. Computation with actual (\ref{fig:BeamPattern-Sol-Th}) and estimated (\ref{fig:BeamPattern-Sol-Cal}) geometries.}
    \label{fig:BeamPattern-Sol}
\end{figure*}

\begin{figure*}
    \centering
    \begin{subfigure}[b]{0.4\textwidth}
        \includegraphics[width=\textwidth]{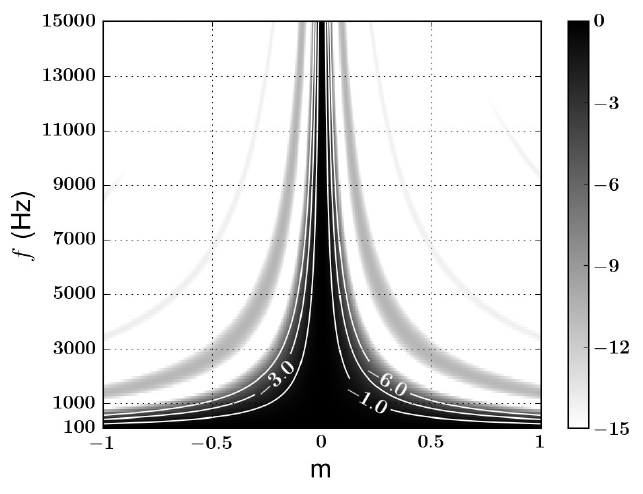}
        \caption{Beamforming reconstructed with the real antenna geometry $\mathbf{X}$, in dB.}
        \label{fig:BeamPattern-Pyr-Th}
    \end{subfigure}
    ~ 
    \begin{subfigure}[b]{0.4\textwidth}
        \includegraphics[width=\textwidth]{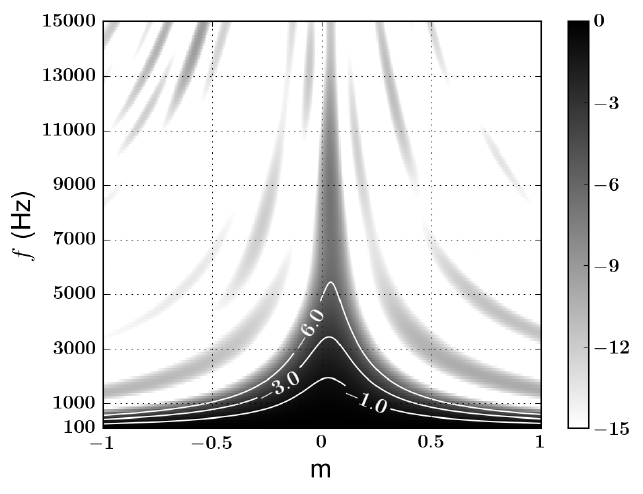}
        \caption{Beamforming reconstructed with the calibrated antenna geometry $\mathbf{\widetilde X}$, in dB.}
        \label{fig:BeamPattern-Pyr-Cal}
    \end{subfigure}
    
    \begin{subfigure}[b]{0.4\textwidth}
        \includegraphics[width=\textwidth]{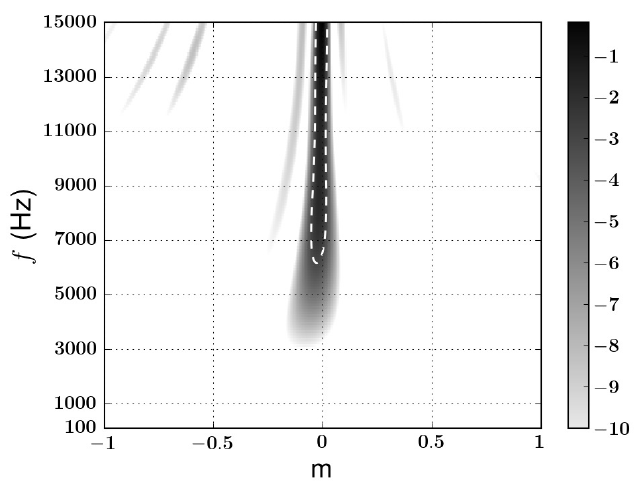}
        \caption{Error on beamforming between \ref{fig:BeamPattern-Pyr-Th} and \ref{fig:BeamPattern-Pyr-Cal}, in dB. $-3$~dB isocontour in white dashed line.}
        \label{fig:BeamPattern-Pyr-Err}
    \end{subfigure}
    \caption{Beamforming in geometrical Near-Field with array 2, with a broadband point source at $5$~m from the array centroid. Computation with actual (\ref{fig:BeamPattern-Pyr-Th}) and estimated (\ref{fig:BeamPattern-Pyr-Cal}) geometries.}
    \label{fig:BeamPattern-Pyr}
\end{figure*}

The results are plotted as a function of frequency, in figures \ref{fig:BeamPattern-Sol} and \ref{fig:BeamPattern-Pyr}. The relative error is also plotted. According to a discrepancy threshold set arbitrarily to $3$~dB, the beamforming shows to perform correctly up to $12$~kHz for the circular array, and $6$~kHz for the pyramid array. These results are consistent with the mean calibration error values of table \ref{tab:error}:
\begin{itemize}
    \item for array 1, $\overline \varepsilon_n = 0.94$~cm; $12$~kHz corresponds to a $2.8$~cm wavelength
    \item for array 2, $\overline \varepsilon_n = 2.05$~cm; $6$~kHz corresponds to a $5.6$~cm wavelength
\end{itemize}

Thus, here the beamformer is globally efficient for wavelengths  decreasing to three times the mean error $\overline \varepsilon_n$. For instance, in the frame of speech localization, both antennas would be  usable. Array 2 (pyramid) though would be unfit at high frequencies for localizing common broadband noise sources. Finally, to quantify the effect of the geometry error at one frequency, figure \ref{fig:BeamPattern-Sol-6kHz} shows the slice of figures \ref{fig:BeamPattern-Sol-Th} and \ref{fig:BeamPattern-Sol-Cal}, \textit{i.e.} it plots the beamformed pressure at frequency $f=6$~kHz for array 1, with the actual and estimated geometries. It shows a $2$~dB decrease of the main lobe, and the maximum position is biased of $2.6$~cm. Also, the highest grating lobe level is $-19$~dB with the actual geometry, and $-17$~dB with the estimated one. Thus the dynamic range, theoretically of $19$~dB, has degraded to $15$~dB. So the localization of the source remains effective at $6$~kHz, but increasing frequency degrades the dynamic range until the main lobe reaches the same magnitude as the grating lobes.

\section{Conclusion}

The present study proposes a robust geometric calibration method microphone arrays of arbitrary shape, in passive diffuse sound field. It is specifically designed for large and wide arrays, \textit{i.e.} having a great number of microphones and an extended spatial range. It relies on the estimation of pairwise distances between microphones, extracted from the measured coherence. Because of the large number of microphones and the spatial extent, the pairwise distances are contaminated by outlying errors. The paper introduces the LRMDS algorithm: it relies both on the knowledge of pairwise distances between close neighbors, and on the unsupervised removal of the outlying errors.

Two experiments were set to validate the process. They show its applicability for 2D and 3D arrays. The measurements were performed in an uncontrolled soundfield. It proves that this calibration method is completely passive, without the need of a specific experimental protocol. The studied soundfield type was in agreement with the isotropic diffuse field model.
In further studies, more complex soundfields could be investigated, such as non-isotropic wave fields, so as to extend the method toward more diversified environments. 

From this frame, the geometry can be known in a few minutes for a further multi-channel process, such as localization or separation of sources, even in three-dimensional scenarios. The present study showed the usability of the method for a Beamforming imaging in geometrical Near-Field: it was evidenced that the localization error is directly linked to a degradation of imaging at high frequencies. The presented results showed an acceptable limit for speech or community noise.

It opens to new possibilities in array applications: very large arrays can be deployed in environments where the diffuse property of the soundfield can be guaranteed. As long as all microphones are connected to several others in a close range, the method remains applicable whatever the total number in the array. In addition to the possibilities offered by MEMS microphones, very large arrays containing more than a thousand of elements can be properly used. Also, further experiments can be investigated beyond an experimental room: it can be deployed for example in an urban environment, where the soundfield comes from spatially scattered uncorrelated sources.
\vfill

\section*{REFERENCES}

\providecommand{\bysame}{\leavevmode\hbox to3em{\hrulefill}\thinspace}
\providecommand{\MR}{\relax\ifhmode\unskip\space\fi MR }
\providecommand{\MRhref}[2]{%
    \href{http://www.ams.org/mathscinet-getitem?mr=#1}{#2}
}
\providecommand{\href}[2]{#2}

\end{document}